# Effects of Beamforming and Antenna Configurations on Mobility in 5G NR

Edgar Ramos, Antonino Orsino
Ericsson Research, Finland
email: {edgar.ramos, antonino.orsino}@ericsson.com

*Abstract*—The future 5G systems are getting closer to be a reality. It is envisioned, indeed, that the rollout of first 5G network will happen around end of 2018 and beginning of 2019. However, there are still a number of issues and problems that have to be faces and new solutions and methods are needed to solve them. Along these lines, the effects that beamforming and antenna configurations may have on the mobility in 5G New Radio (NR) is still unclear. In fact, with the use of directive antennas and high frequencies (e.g., above 10 GHz), in order to meet the stringent requirements of 5G (e.g., support of 500km/h) it is crucial to understand how the envisioned 5G NR antenna configurations may impact mobility (and thus handovers). In this article, first we will briefly survey mobility enhancements and solution currently under discussion in 3GPP Rel. 15. In particular, we focus our analysis on the physical layer signals involved in the measurement reporting and the new radio measurement model used in 5G NR to filter the multiple beams typical of directive antenna with a large number of antenna elements. Finally, the critical aspect of mobility identified in the previous sections will be analyzed in more details through the obtained results of an extensive system-level evaluation analysis.

*Keywords—5G; NR; Mobility; Handover; Beamforming; Antenna configuration, Measurements*

## I. INTRODUCTION

THE mobile networks success has relied in the intrinsic characteristic of their overall availability and apparently seamless support for continuous connectivity. The 5th generation of networks are expected to manage even more challenging requirements than the previous generation with an increased performance and improved end-user experience [1]. In the mobile radio network, there is consensus in the industry that the main feature contributing to the bursting of capacity and performance is beamforming [2,3]. Beamforming is required to overcome the limitations introduced by the deployment of the radio network in high frequencies (> 2 GHz) in the search of more bandwidth to translate in higher data rates. The operation of radio sites based in beamforming requires a change in the paradigms of mobility than has been used so far by LTE and UMTS. The complexity of mobility is expected to rise due to the extra-dimension introduced by the beam management in addition to densification required to provide coverage in higher frequencies [1]. In addition to the data rates, the reliability and latency are also expected to be improved to such extend that latency critical processes requiring very low error probabilities could make use of the 5G connections to communicate between system components. This translate in a virtual zero interruption of data during handover and a very low probability of total handover failure [4].

According to such stringent requirements, in this paper we want to shed light about the issue and challenges that have to be faced in the next-to-come 5G systems. In doing this, we provide a general overview about proposed solutions that, at the time we are writing, are under discussion in the 5G New Radio (NR) 3GPP Rel. 15 standardization. In particular, our study is mainly focused on two mobility aspects: (i) the physical layer signal to be reported and taken into account for handover and (ii) how the different antenna configuration that will be most likely used in 5G NR will affect the mobility (handover) procedures.

The first aspect is referred to the cell- and beam-quality measurement reporting that, if done in an efficient wat, it helps to increase the handover robustness thus reducing the handover failure rate. The second aspect, instead, aims at identify which are the aspects to be improved in order to reduce the radio link failure during the handover procedures. For instance, one possible solution could be the introduction of additional beam information (e.g., beam-specific identifier towards the source or neighbor cells).

The remainder of this paper can be summarized as follows. In Section II are surveyed general aspects about new complexity concerning mobility in 5G NR. Possible issues and solution are, instead, described in Section III whereas the new measurement model for 5G NR is illustrated in Section IV. Finally, in Section V we discuss the results obtained with our system-level evaluation campaign with the final remark provided in Section VI.

## II. 5G MOBILITY COMPLEXITY

There is a trend in the industry [5] to handle the mobility complexity by centralizing the mobility management either by concentrating the actual control or by aggregating several end points of transmission and treat them as a big composite area. The area is covered by a set of basically transmission points (TPs), such as distributed antennas or MRUs connected to one central processing unit. This would allow to control the mobility of one area without needing to exchange control signaling between the network and the terminal, which is recognized as one of the main causes for handover failure [5].

Another challenge is the different requirements of mobility during the different activity states of the terminals. Terminals in IDLE mode and power saving states do not need and are not able to provide updates and signal the radio network frequently. In the other hand, for terminals that are in CONNECTED mode and therefore actively exchanging information with the network, is critical to maintain their connection and are able to easily update the network with measurements and any type of configured reports.

The 3GPP standardization effort for 5G called NR (New Radio) has already defined two physical radio signals that can be used alternatively or conjunctly to manage mobility. The Reference Signals (RS) properties haven't fully been specified at the time of writing this article, but according to the already agreed features and the discussions between the different parties involved, the main characteristics are possible to be inferred. The first RS is a similar signal to the PSS/SSS used in LTE. Similarly, as the LTE counterpart, it counts with a synchronization component and for the sake of this article we name it NR-SS (New Radio Sync Signal). This RS is intended to be utilized specially for IDLE mode mobility, which means that the terminals use such signal to differentiate different coverage areas (NR Cells) and proceed to acquire the configuration broadcasted in the NR-Cell and to start an initial access to transmit messages to the network, such as, measurement updates, user plane data or connection setup signaling. The terminals are able to find the broadcast channel by means of detecting and synchronizing to the NR-SS.

The second RS defined has been denoted as CSI-RS (Channel State Information Reference Signal). This RS is expected to be very similar to the corresponding LTE signal with the same name, but with some additional enhancements. The CSI-RS are expected to be used in scenarios with NR-Cells comprised by multiple transmission points, deployments at high frequencies or with large number of antennas, centralized mobility control, and terminal requiring services depending on flawless mobility performance.

The main dilemma for the NR design of the mobility signals is the tradeoff between an energy efficient signal that should also provide a sufficient good performance for fast moving UEs (up to 500Km/h) and works also for challenging propagation environments with high probability of blockage (frequencies > ~30Ghz). Furthermore, it is desirable that the properties of the signals allow a fast and efficient searching function, especially with the increase in the searching space given the NR bandwidth targets (up to 1 GHz and more). For a terminal in IDLE mode would mean a relatively blind search, since it does not count with any other configuration data than the statically defined by the standard or the one provided by the SIM card. Meanwhile a flexible solution, energy and spectrum efficient, requires that the terminal is configured with, at least, restricted searching spaces (in time and frequency domains) to be able to locate the reference signals.

The mobility solution involves considering all these requirements that seems conflicting at times and which level of importance and priority are highly depending on the actual deployment and the type and mobility patterns of the traffic present. A macro deployment with few antennas and operating in frequencies similar to LTE is more likely to benefit from the simplicity of a single RS, beamformed or not, operating in a similar fashion as the LTE mobility does. The RS would cover the NR-Cell area and the terminals would report a "cell quality" based in RS measurements, for example, Reference Signal Received Power (RSRP) or Reference Signal Received Quality (RSRQ). In the case of RS beamforming, a periodic beam sweeping of the signal is used to cover the whole NR-Cell area. The details of beamforming, such as, analog vs digital, number and special distribution of beams, as well as the difference in capabilities between terminals and base stations (antennas, bandwidth handling, frequency bands, etc.) complicates further the mobility procedures and configuration.

More complicated deployments are those handling multiple TPs, higher frequencies with high densification, targeting indoor systems or mixed indoor/outdoor coverage, high-rise urbans, or when performance critical (latency and reliability) applications are offered as a service.

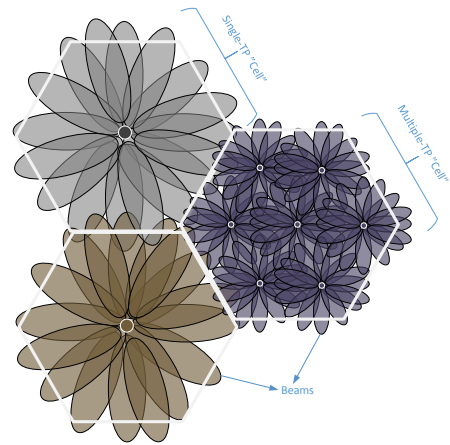

Fig. 1. Example of deployment with a grid of beams and one site with multiple TPs.

The mobility inside the whole coverage area of a NR-Cell (intra-cell mobility) is transparent to the terminal and is realized by means of beam management.. The mobility between NR-Cells (inter-cell mobility) requires to identify what are the mobility candidate NR-Cells and decide the target(s) after a handover procedure, including handover preparation, configuration, and execution (and in some cases contingency after failure). To address the inter-cell mobility, it has been defined two different approaches which the industry has labeled as uplink based and downlink based mobility. The first one uses reference signals transmitted by the terminals that allows the network to identify the position and quality of the connection of a particular terminal [6], this approach is out of the scope of this article. For the second approach, the terminals are required to send measurement reports to the network to identify and quantize the quality of the reference signals transmitted by the NR-Cells.

III. INTER-CELL MOBILITY IN 5G

The mobility between NR-Cells is one of the focus areas of interest on the mobility topic. In first place because is a heavy procedure that is at the core of the capability of the mobile radio networks, it requires a considerable amount of critical signaling between the network and the terminal which usually is carried out with the worst channel conditions at the cell borders. Therefore, the risk of failure is an ever-present possibility during each handover and the system should be configured to minimize such problems. Also, the correct identification of the best target cell and to proceed the preparations for the handover before it is applied has a direct impact in the performance. In the case of multi-hop networks and mobile relays (moving TPs in contrast to regular fix TPs) the network should be able to additionally take in account the routing and use of supporting resources from the donor (or anchor) to service the terminals connected to the serving TP. These last deployments type are also out to the scope of this article and we will focus in the fixed TPs deployments.

In a 5G-NR system, the heterogeneous characteristics of beamforming makes the assessment of the cell borders and the need of a handover difficult . The beamforming gain is not the only differentiation factor compared to a LTE network. Other aspects are the actual shape and characteristics of the beam (azimuth angle, wide, overlap with other beams, transmission periodicity, etc.), as well as the characteristics of the receiver. This would have an impact in what are the absolute and relative values reported by terminals in the measurements. In fact, the use of high frequencies (i.e., up to 100 GHz) led to a more complex granularity for the user that need to perform signal measurements not only for each TP, but also for each beam within belonging to a given TP. Along this line, in the last 3GPP meetings regarding the standardization of 5G NR has been agreed that the terminal (i.e., the UE), in addition to the standard measurements typical of LTE systems, has to estimate the signal quality over "N" best beams (either towards the source and neighbors cells/TPs) and send a report to the network. Then, based on the received UE measurements, the network decides when to trigger and how to manage the handover.

Another point that cannot be overestimate, is that one related to the type of measurements that may be useful to optimize the handover procedure thus guarantee a handoff and session continuity close to 0 ms. The UE, indeed, may apply different approaches in performing the measurements. Along this direction, the UE may provide an "average" (over the time) signal quality along the selected best beams whereas another UE may provide the "instantaneous" signal quality. Of course, this leads to a heterogeneity of possible measurements that has to be harmonized in the close future.

Although handover procedures and what has to be measured is still under discussion, the main conclusion for the time being is that NR we will have much more variables and complexity w.r.t. what we had so far for LTE. Just to give an example, if Fig. 2 is shown the signal strength measured in 1000 positions of a macro deployment operating at 28 GHz for different antenna configurations of the network base stations.

As we can observe, the signal threshold (i.e., identified by the area between the green and light blue color) that a UE most likely will use to trigger the handover command to the source TP is pretty different based on the antenna configuration used.

In a scenario where TP belonging to the same cell may have different antenna configuration, this translates in a high number of ping pong effects and handover failures since the UE is not aware of the overall configuration of the deployed network. Therefore, new methods and solutions for handle the mobility (i.e., on the handover and measurement side) are needed for facing the challenging requirements of next-to-come 5G systems.

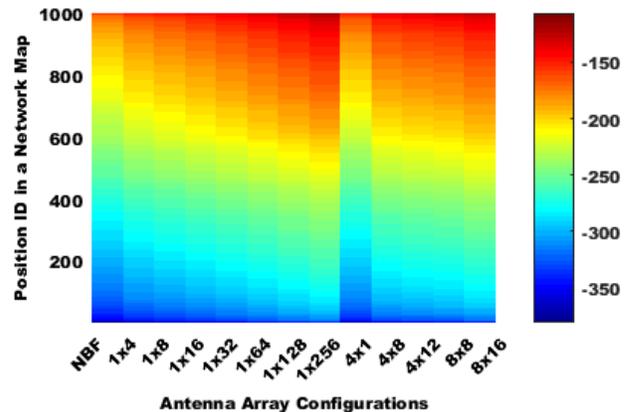

Fig. 2. Signal strength (dBm) for different antenna configurations and Not Beam Forming (NBF) at 28 GHz for 1000 positions of a macro deployment

## IV. MEASUREMENT MODEL IN 5G NR

One of the main differences of 5G NR with respect to the legacy LTE networks regarding the measurements model is that this one has to handle different antenna configuration where narrowed beams are used. This is particularly evident when using millimeter frequencies (e.g., 28 GHz). In such a case, the measurements are not done over the typical three sectors deployed in a microcell site but should be performed to a grid of beams that may be quite large (i.e., as shown in Fig. 1). Along these lines, in past 3GPP meeting regarding NR have been done a series of enhancements in order to take into account the presence of the beams. In fact, according to the new measurement model for 5G NR [7], a UE that is in RRC_CONNECTED state measures multiple beams (or if it not possible, at least one) belonging to a cell and the measurements results involving metrics such as RSRP, RSRQ, or SINR. are averaged to derive the overall cell quality. In doing so, once the detection of all the possible beams has been complete, the network configures the UE to consider only a subset of the detected beams (the *N* best beams) that are above an absolute threshold.

Further, as for LTE, the filtering takes place at two different levels. One is at the physical layer where beam quality is derived and the other one is at RRC level where the overall cell quality from multiple beams is achieved. We note that the measurement reports send periodically (or not) by the UE to the network may contain the measurement results of the *N* best beams (or a subset of them) if the UE is configured to do so by the gNB.

The new RRM model used in 5G NR is shown in Fig. 3 where the different steps to be performed during the L1 and L3 filtering are illustrated. It is worth noticing, however, that some of the procedures are specified in the 3GPP NR specifications (e.g., in 3GPP TS 38.331) but others are left to the UE/network implementation e.g., how and when the UE exactly performs the required measurements. Nevertheless, since the RRM model of 5G NR is out of the scope of this work, for further information about how it works and what are the specific functionalities, the reader can refer to the 3GPP TS 38.330 specification for further details.

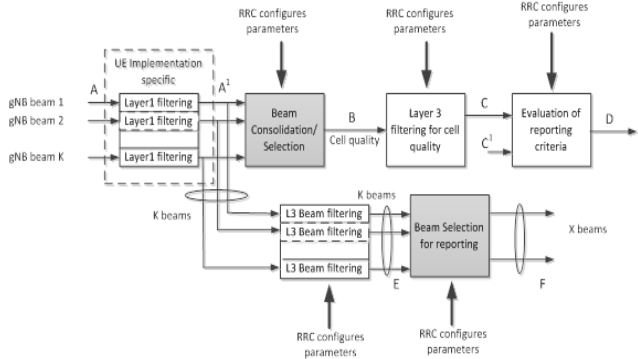

Fig. 3. New RRM model for 5G NR [7]

## V. PERFORMANCE EVALUATION

In order to shed light on the impact that different antenna configurations (and thus number of beams) may have on the handover procedures and mobility behavior in general, an extensive performance evaluation has been conducted with the help of a proprietary system-level simulator.

In the scenario considered, three macro base station sites with tri-sectorial antennas are deployed with an inter-site distance of about 200m. The radio access technology used in the simulation is the 5G NR over a frequency of 3.5 GHz with 40 MHz bandwidth. The environments simulated are both indoor and outdoor where 5 UEs are deployed within the area of interest and move with a speed that varies from 3 km/h to 30 km/h. The network is modeled with an FTP download session where a file of 200Mbits is divided in 10 chunks of 20Mbits each one of them download with an interval of 1.5s. A grid of beams is deployed according to the antenna configuration, and CSI-RS are used for the mobility measurements. The antenna configurations on the gNB investigated varies from 16 to 128 antenna elements ((i.e., 16, 32, 64, and 128) and the metrics evaluated in our performance analysis are: (i) the *Serving RSRP*, that is the RSRP of the beam selected by the UE to establish a transmission, (ii) the *Best RSRP*, that is the RSRP of the best beam among those sensed by the UE, and the (iii) *Delta RSRP*, that is the offset between the RSRPs of the best beams sensed by the UE and the network (while the network is aware of all the beams available, the UE is configured to sense only a part of them). Further, the remaining simulation parameters can be found in Table 1.

Table 1. Main simulation parameters

| Parameter | Value |
|---|---|
| Number of macro sites | 3 |
| Number of sectors per site | 3 |
| Height of BS | 25m |
| Height of UE | 1.5m |
| BS Tx power | 43 dBm |
| UE Tx power | 23 dBm |
| Number of UEs | 15 |
| Speed of UEs | [3-30] km/h |
| Inter-site distance | 500m |
| NR Frequency | 3.5 GHz |
| NR bandwidth | 40 MHz |
| Antenna elements | [16, 32, 64, 128] |
| Pathloss model | 3GPP 3D SCM |
| Slow fading | 3D SCM slow fading UMa |

### A. Obtained Results

In the first results we present in our analysis we provide a comparison of the Serving RSRP and Best RSRP (in reference to the beams detected by the UE) when considering the outdoor and indoor scenarios. In particular, as we can see from Fig. 4(a) and Fig. 4(b) the RSRP (and thus the beam) chosen by the UE lead to a higher signal strength in case of outdoor scenario. The motivation behind this is that on the pathloss propagation model (i.e., the #D SCM) additional losses (e.g., walls) are added when users are indoor. Further, in Fig. 4 when comparing the Serving RSRP and Best RSRP in general, we observe that not always the UE select the best available beam among the set of *N* best beams measured. However, this behavior pretty much depends on the grid of beams available, the mobility of the users (and thus their position over time) and the L1 and L3 filtering parameters on the RRM model. Nevertheless, all at all it is worth noticing that when the number of antenna elements is high (e.g., 128) the signal strength with which the UE is served improves. This is mainly due to the spatially separation of the beams that helps to reduce the interference. In fact, UEs that are relatively close to each other may be eventually be served with two different beams

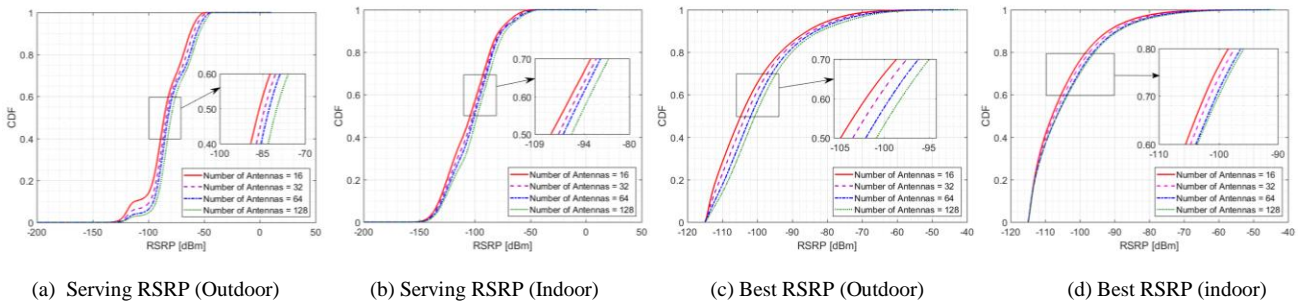

(a) Serving RSRP (Outdoor)      (b) Serving RSRP (Indoor)      (c) Best RSRP (Outdoor)      (d) Best RSRP (indoor)

Fig. 4. Serving and Best RSRP with different grid of beams and one site with multiple TPs.

and thus not interfering strongly to each other.

Thinking about how this may affect mobility and handovers in NR, it is easy to say that when using a large number of antenna elements, it is possible to have higher (and in certain condition more stable) signal strength and, at the same time, higher data rates and low latencies. The drawback is that the parameters within the measurement model have to be tuned accordingly in order to lower ping pong effect and unnecessary handovers. If fact, in such a case measurements reporting, the way how metrics are evaluated (i.e., average, absolute, or instantaneous value) play a fundamental role to reach the stringent requirements of next to come 5G systems.

The second result we want to highlight from our conducted system-level simulations, is the Delta RSRP for the outdoor and indoor case (i.e., please refer to Fig. 5). This value shows the misalignment between the absolute best beam out of the all beams generated by the antennas on the BS side and the best beam among those one selected by the UE. We recall that the network configures the UE (through the RRM model) to select a set of $N$ best beams thus all the beams generated by the antenna are filtered to reduce complexity on the terminal. (i.e., UE). The results obtained in Fig. 5, show that the best beam available at the UE is not always the best of the overall grid of beams irradiated by the BS. Therefore, the showed behavior is particularly important when considering condition for triggering the handover and filtering to be applied by the UE in the RRM model. In fact, this highlight that if we want to achieve the 0ms handover interruption time settled by the ITU and 3GPP, still some work has to be done and further enhancements need to be proposed. Along these lines, what is clear is that the usage of beams and antennas with a larger number of elements bring a greater complexity with respect to LTE that has to be taken into account by both the UE and BS. Even if on one side beamforming provide very high directivity and thus high data-rate and low latency, on the other side the effect on the handover failure or ping-pong handover may be relevant.

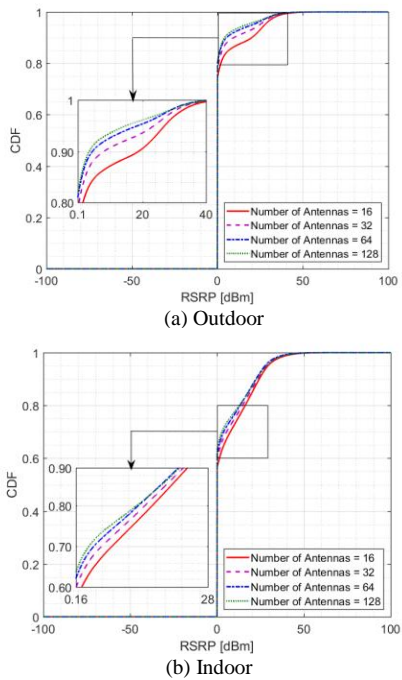

Fig. 5. Delta RSRP with different grid of beams and one site with multiple TPs.

## VI. CONCLUDING REMARKS

In this paper we shed light about the issues and challenges that have to be faced in the next-to-come 5G systems for what concern the effects of beamforming and different antenna configuration on the mobility of 5G NR. Our analysis focused on the physical layer measurements that typically are reported by the UE to the network and how those measurements are performed when considering directive antennas. Then, the critical aspects of mobility identified and described have been analyzed in more details through an extensive system-level evaluation analysis. Obtained results showed that when using a large number of antenna elements (e.g., 128 elements) it is possible to achieve higher data rate and low latency even if the tuning of parameters on the measurements model become more challenging in order to lower the ping pong effect and unnecessary handovers.